\documentclass[aps,prb,twocolumn,showpacs,psfig,superscriptaddress,longbibliography]{revtex4-1}

\usepackage{textcomp}
\usepackage{times}
\usepackage{graphicx}
\usepackage{float}
\usepackage{latexsym,amsmath,amssymb,bm,euscript}
\usepackage{color}
\usepackage{subfigure}
\usepackage{epstopdf}
\usepackage[colorlinks=true,linkcolor=blue,citecolor=blue]{hyperref}
\usepackage{hyperref}
\usepackage{soul}
\usepackage[normalem]{ulem}
\usepackage{mathrsfs}
\usepackage{amsmath}
\usepackage{lettrine}
\usepackage{xspace}
\usepackage{textcomp}
\usepackage{xr}
\usepackage{appendix}
\usepackage{verbatim}

\DeclareGraphicsExtensions{.pdf, .png, .jpg, .eps}

\begin{document}

\title{Field-induced phase transitions and quantum criticality in a honeycomb antiferromagnet Na$_3$Co$_2$SbO$_6$}

\author{Ze~Hu}
\thanks{These authors contributed equally to this study.}
\affiliation{Department of Physics and Beijing Key Laboratory of
Opto-electronic Functional Materials $\&$ Micro-nano Devices, Renmin
University of China, Beijing, 100872, China}

\author{Yue~Chen}
\thanks{These authors contributed equally to this study.}
\affiliation{International Center for Quantum Materials, School of Physics,
Peking University, Beijing 100871, China}

\author{Yi~Cui}
\affiliation{Department of Physics and Beijing Key Laboratory of
Opto-electronic Functional Materials $\&$ Micro-nano Devices, Renmin
University of China, Beijing, 100872, China}
\affiliation{Key Laboratory of Quantum State Construction and Manipulation (Ministry of Education),
Renmin University of China, Beijing, 100872, China}

\author{Shuo~Li}
\affiliation{Department of Physics and Beijing Key Laboratory of
Opto-electronic Functional Materials $\&$ Micro-nano Devices, Renmin
University of China, Beijing, 100872, China}

\author{Cong~Li}
\affiliation{Department of Physics and Beijing Key Laboratory of
Opto-electronic Functional Materials $\&$ Micro-nano Devices, Renmin
University of China, Beijing, 100872, China}

\author{Xiaoyu~Xu}
\affiliation{Department of Physics and Beijing Key Laboratory of
Opto-electronic Functional Materials $\&$ Micro-nano Devices, Renmin
University of China, Beijing, 100872, China}

\author{Ying~Chen}
\affiliation{Department of Physics and Beijing Key Laboratory of
Opto-electronic Functional Materials $\&$ Micro-nano Devices, Renmin
University of China, Beijing, 100872, China}

\author{Xintong~Li}
\affiliation{International Center for Quantum Materials, School of Physics,
Peking University, Beijing 100871, China}

\author{Yuchen~Gu}
\affiliation{International Center for Quantum Materials, School of Physics,
Peking University, Beijing 100871, China}

\author{Rong~Yu}
\affiliation{Department of Physics and Beijing Key Laboratory of
Opto-electronic Functional Materials $\&$ Micro-nano Devices, Renmin
University of China, Beijing, 100872, China}
\affiliation{Key Laboratory of Quantum State Construction and Manipulation (Ministry of Education),
Renmin University of China, Beijing, 100872, China}

\author{Rui~Zhou}
\affiliation{Beijing National Laboratory for Condensed Matter Physics and Institute of Physics,
Chinese Academy of Sciences, Beijing, 100190, China}

\author{Yuan~Li}
\email{yuan.li@pku.edu.cn}
\affiliation{International Center for Quantum Materials, School of Physics,
Peking University, Beijing 100871, China}
\affiliation{Collaborative Innovation Center of Quantum Matter, Beijing 100871,
China}

\author{Weiqiang~Yu}
\email{wqyu\_phy@ruc.edu.cn}
\affiliation{Department of Physics and Beijing Key Laboratory of
Opto-electronic Functional Materials $\&$ Micro-nano Devices, Renmin
University of China, Beijing, 100872, China}
\affiliation{Key Laboratory of Quantum State Construction and Manipulation (Ministry of Education),
Renmin University of China, Beijing, 100872, China}

%\date{\today}

\begin{abstract}

We performed $^{23}$Na NMR measurements on a single-domain
crystal of the Kitaev material Na$_3$Co$_2$SbO$_6$,
with magnetic field applied along the crystalline $a$ axis.
A positive Curie-Weiss constant is obtained from the NMR Knight shift,
which suggests the existence of ferromagnetic exchange couplings.
The antiferromagnetic ordering is found to be suppressed
at a field of 1.9~T.
Inside the ordered phase, our data reveal two additional phase
transitions. At 1.9~T, the spin-lattice relaxation rate $1/^{23}T_1$
establishes a quantum critical behavior at high temperatures.
However, at low temperatures, a gapped behavior
is observed at the ``critical" field,
which suggests a weakly first-order transition
instead and a possible field-induced quantum spin liquid.
Our results reveal complex microscopic interactions
in the system, which may help to search for possible
quantum spin liquids.

\end{abstract}

\maketitle

\section{{\label{sintro}} Introduction}
Experimental search for quantum spin liquids (QSLs),
representing a disordered phase beyond Landau's paradigm, containing novel properties such
as fractional excitations and long-range entanglement, and promising for
unconventional superconductivity and quantum computation, has been a
heated frontier in condensed matter physics~\cite{Anderson_MRB_1973,Balents_Nat_2010,
Anderson_science_1987,A.Yu.Kitaev_AoP_2003,Nayak_RMP_2008}.
In antiferromagnetic (AFM) systems with triangular, kagome, and pyrochlore lattice structures,
strong geometric frustration may be sufficient to suppress magnetic ordering and
lead to QSLs.  However, it is highly debated if QSLs are established in real materials,
because site or anti-site disorder, which strongly affects the nature of the ground states,
has been frequently reported~\cite{Shimizu_PRL_2003,Han_Nat_2012,Gardner_RMP_2010}.

In parallel, the Kitaev model in the honeycomb lattice, which contains exchange
frustration among neighboring bonds with orthogonal Ising-type couplings,
is a rare 2D case where the QSL is an exact ground state with Majorana
and photonic gauge excitations~\cite{Kitaev_AP_2006,Motome_JPSJ_2020,Takagi_Na_2019}.
Theoretical studies also predict that AFM Kitaev model
carries either Z$_2$ or U(1) gauge field~\cite{Hickey_NC_2019,Hickey_PRR_2020,Kaib_PRB_2019},
whereas ferromagnetic (FM) Kitaev model only hosts Z$_2$ type.
Until recently, it is proposed that some 4d or 5d
transition metal ions with hexagonal lattice structures,
such as $A$$_2$IrO$_3$ ($A$=Li, Na)~\cite{Choi_PRL_2012,Sae_NP_2015,Kimchi_PRB_2011,Kim_PRX_2020}  and $\alpha$-RuCl$_3$~\cite{Plumb_PRB_2014,Kubota_PRB_2015,Baek_PRL_2017,Do_NP_2017,Banerjee_NM_2016,
Stone_Sci_2017,Zheng_PRL_2017,Banerjee_QM_2018,Sears_NP_2020,Kasahara_Nat_2018,YBKim_PRB_2018}, may contain
Kitaev interactions arising from strong spin-orbit coupling and bond symmetries~\cite{Jackeli_PRL_2009,Rau_PRL_2014}.
To be more interesting, hexagonal cobaltates,
such as Na$_2$Co$_2$TeO$_6$, Na$_3$Co$_2$SbO$_6$, and BaCo$_2$(AsO$_4$)$_2$, are
proposed to be a 3d transition metal class of QSL candidate materials,
due to interplay among trigonal crystal field, charge transfer,
spin-orbit coupling, and Coulomb interactions~\cite{Liu_PRB_2018,Sano_PRB_2018,Liu_PRL_2020,Kim_JPCM_2021,Liu_JMPB_2021}.

However, in all of these Kitaev materails, AFM ground states are
usually observed~\cite{Bera_PRB_2017,Wong_JSSC_2016,Yan_PRM_2019,Das_PRB_2021,Viciu_JSSC_2007,Lefran_PRB_2016,Yao_PRB_2020,Xiao_CGD_2019}.
It is then proposed that these systems should be described by a combined $H$-$K$-$\varGamma$ model, where $H$
stands for isotropic Heisenberg, $K$ for Kitaev, and $\varGamma$ for off-diagonal spin couplings,
and QSL may only survive in narrow, barely reachable parameter spaces~\cite{Luo_QM_2021,Takagi_Na_2019,Gotfryd_PRB_2017,Maksimov_PRX_2019}.

\begin{figure}[t]
\includegraphics[width=8.5cm]{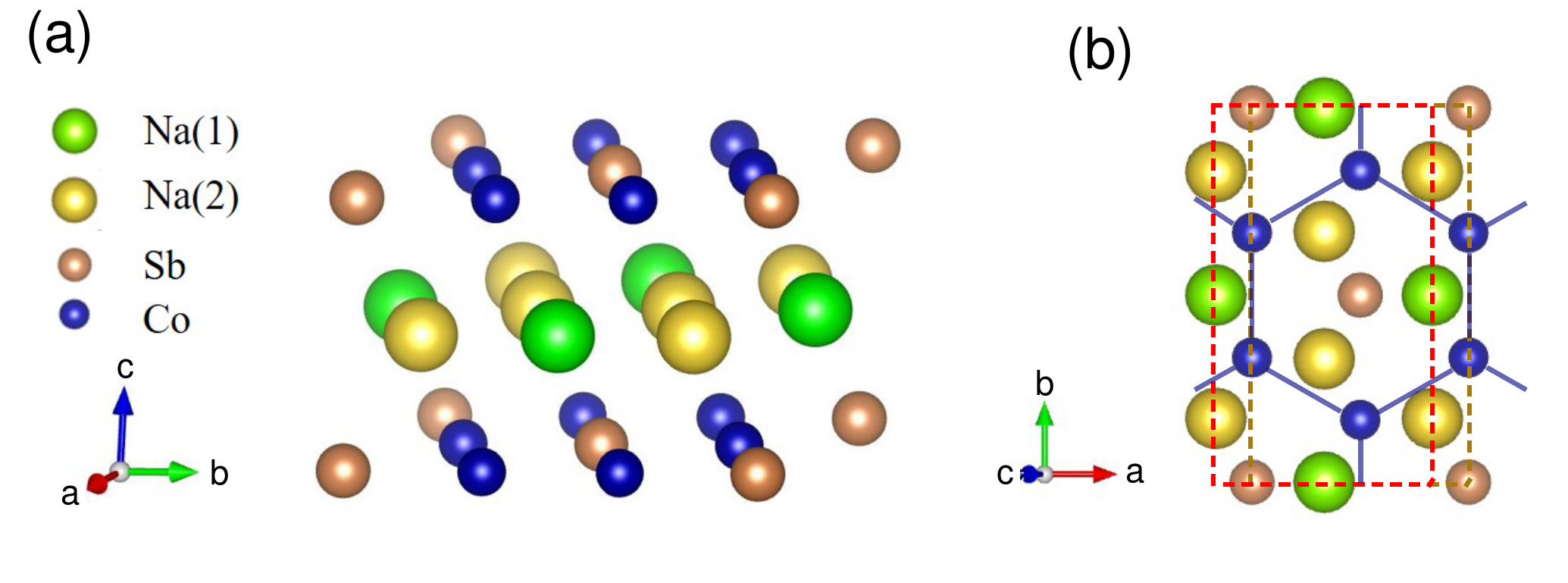}
\caption{\label{fstruc}
  {\bf Lattice structure of Na$_3$Co$_2$SbO$_6$.}
  (a)~Side view of one unit cell of the compound,
  where magnetic Co$^{2+}$ ions form the layered regular honeycomb lattice,
  with Sb$^{5+}$ ions located at the center of each hexagon.
  Oxygen atoms in the edge-shared CoO$_6$ octahedra are omitted for simplicity.
  Na$^{+}$ ions separate the magnetic layers with
  two types of inequivalent sites, labeled as Na(1) and Na(2).
  (b) Top view of one half of unit cell along the $c$ axis. The red and brown dotted lines
  depict the in-plane boundary which contains Na$^{+}$ and Co$^{2+}$ ions respectively.
  Na(1) and Na(2) differ slightly by their distances to neighboring Co-Co bonds.}
\end{figure}

Na$_3$Co$_2$SbO$_6$, as shown in Fig.~\ref{fstruc}(a),
contains edge-shared CoO$_6$ octahedra.  Co$^{2+}$ ions, with effective spin-1/2,
form a layered honeycomb lattice, with SbO$_6$ octahedra located at the center in the same layer.
The exchange couplings among Co$^{2+}$ are bridged by both edge-shared oxygen (through $p$ orbitals)
and Sb (through $d$ orbitals) atoms.
Dominate FM Kitaev interactions have been proposed theoretically~\cite{Liu_PRL_2020}, but yet to be
proved experimentally.
At zero field, the compound is ordered below $T_{\rm N}\approx$~6.6~K~\cite{LiYuan_PRX_2022},
with an AFM pattern and a propagation vector $K$ = (1/2, 1/2, 0)~\cite{Wong_JSSC_2016,Yan_PRM_2019}.
The magnetic structure of Na$_3$Co$_2$SbO$_6$, as well as Na$_2$Co$_2$TeO$_6$, may not follow the simple zigzag
pattern~\cite{Chen_PRB_2021,Yao_PRR_2023,Lee_PRB_2021,GuYuchen_arXiv_2023}
and varies with magnetic field~\cite{Yao_PRB_2020,LiYuan_PRX_2022}.
Magnetic excitations in Na$_3$Co$_2$SbO$_6$, measured by inelastic neutron scattering (INS),
$\mu$SR and NMR measurements, reveal strong quantum fluctuations and possible
coexistence of $H$, $K$ and $\varGamma$ exchange
couplings~\cite{Kim_JPCM_2022, Lin_NC_2021,Songvilay_PRB_2020,Yao_PRL_2022, Miao_arXiv_2023,Vavilova_PRB_2023}.
However, an easy-plane XXZ model is also suggested by a recent study~\cite{GuYuchen_arXiv_2023}.
To further address these issues and search for QSL, more spectroscopic studies through tuning
are highly desired~\cite{Zheng_PRL_2017,Jia_CPL_2020,Liu_PRL_2018}.

In this work, we report $^{23}$Na NMR studies
on a high quality, twin-free single crystal of Na$_3$Co$_2$SbO$_6$.
We identify a positive Curie-Weiss constant from the NMR Knight shift
in the paramagnetic (PM) phase.
The magnetic orderings are confirmed by the NMR spectra and
the transition temperatures are resolved by the
NMR spin-lattice relaxation rates $1/^{23}T_1$.
With field applied along the crystalline $a$ axis,
two additional magnetic phase transitions are found in the ordered phase,
as shown by the detailed phase diagram (see Fig.~\ref{fqcr}).

Low-energy spin dynamics, revealed by
the spin-lattice relaxation rates, $1/^{23}T_1$, demonstrates a
field-suppression of magnetic order. In particular,
a funnel shape in the color
map of $1/^{23}T_1T$ is established at a field of 1.9~T, which is a
clear evidence for quantum criticality
in the paramagnetic phase.
However, the low-temperature $1/^{23}T_1$ reveals a possible weakly first-order quantum
phase transition at 1.9~T, and a possible QSL at higher fields.
Our study reveals the complexity of phases and magnetic exchange couplings in the system,
and promotes further investigations on field-induced phases, such as QSLs.

\section{{\label{smethod}} Materials and Techniques}

High quality single crystals were grown by the chemical vapor transport method~\cite{Xiao_CGD_2019}.
A single domain (twin-free) sample was selected for the current study.
The usage of single domain crystal allows to induce a single magnetic phase
with one field orientation, which also simplifies the NMR spectra.
A recent low-temperature magnetization study reveals complicated phase transitions
with different in-plane field orientation~\cite{LiYuan_PRX_2022}.
In this study, the field is applied along the crystalline $a$ axis,
which results in a larger critical field~\cite{LiYuan_PRX_2022} and benefits NMR measurements.
The magnetic field is calibrated by the  $^{63}$Cu resonance frequency
from the NMR tank coil.

A top tuning circuit was used to cover a wide frequency range with fields from 1~T to 10~T.
The $^{23}$Na NMR spectra were collected by the standard spin-echo technique.
For verification, we also performed $dc$ magnetization measurements
at selected temperatures in a Magnetic Property Measurement System (MPMS),
with the same field orientation.

$^{23}$Na is an isotope with spin $I$~=~3/2 and Zeeman factor $^{23}\gamma$~~=~11.262~MHz/T.
The NMR Knight shift $^{23}K_n$ is calculated by $K_n = (f-\gamma H)/\gamma H$,
where $f$ is the peak frequency of the center NMR lines.
The NMR spin-lattice relaxation rate $1/^{23}T_1$ is obtained by the inversion-recovery method,
with spin recovery curve fitted to the exponential function for spin-3/2 nuclei,
$I(t) = a-b[e^{-(t/T_1)^\beta}+9e^{-(6t/T_1)^\beta}]$, where $\beta$
is the stretching factor.

\section{{\label{sspec} NMR spectra}}

\begin{figure*}[t]
\includegraphics[width=15cm]{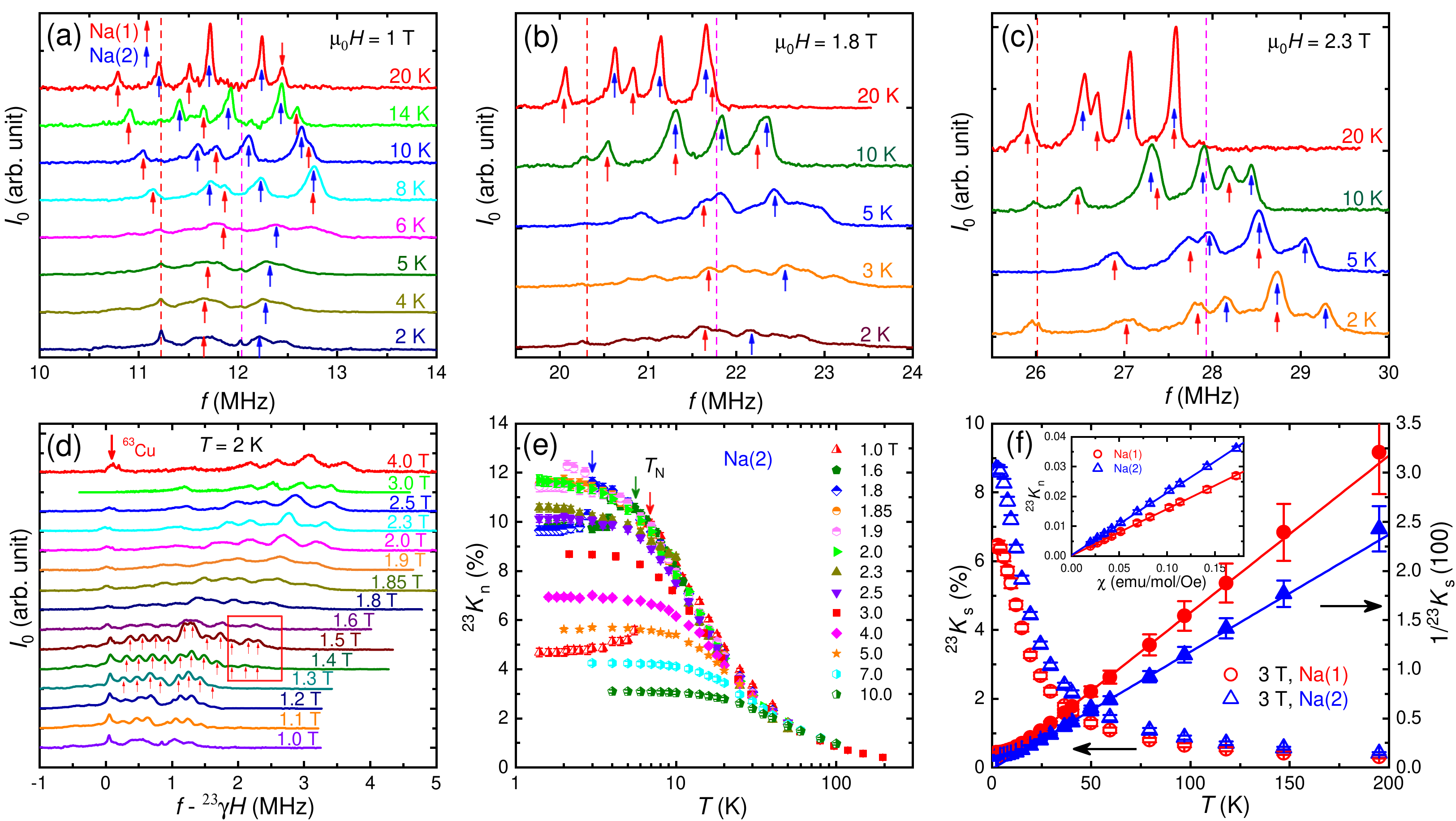}
\caption{\label{fspec}
{\bf NMR spectra.}
  (a)-(c) $^{23}$Na spectra measured at different temperatures, with representative fields of
  well below (1.0~T), close to (1.8~T), and above (2.3~T) the field-induced quantum phase transition.
  Red and blue arrows label the Na(1) and Na(2) spectra respectively. The orange (magenta) dash lines
  denote the position of $^{63}$Cu ($^{65}$Cu) lines from the NMR coil.
  (d) Full $^{23}$Na NMR spectra taken at a constant temperature of 2~K with increasing fields.
  The red rectangle denotes emerged peaks in the 1/3-AFM phase at fields of 1.4~T and above.
  The up-arrows mark all resolvable peaks across the phase transition.
  (e) Knight shift $^{23}K_n$ as functions of temperatures measured at different fields.
  (f) Knight shift $^{23}K_s$ (left scale) and $1/^{23}K_s$ (right scale) plotted as functions of temperatures, measured on both Na(1) and Na(2) at 3~T.
  Solid lines are linear fits to $1/^{23}K_s$ with temperature from 25 to 200~K.
  Inset: $^{23}K_n$-$\chi$ plot with temperature from 25 to 200~K. Solid lines are linear fits to obtain chemical shift $K_C$ and $A_{hf}$.
}
\end{figure*}

In order to investigate local magnetic properties,
the NMR spectra were collected at various magnetic fields and temperatures,
with frequencies relative to $^{23}{\gamma}H$, as depicted in Fig.~\ref{fspec}(a)-(c).
At $T=$ 20~K, six $^{23}$Na NMR lines are identified,
which correspond to two types of interlayer Na atoms, Na(1) and Na(2) as
shown in Fig.~\ref{fstruc}(b), and each has one center
peak and two satellite peaks.
With an occupancy ratio of $N_{\rm 1}$$:$$N_{\rm 2}$=$1$:$2$ in the lattice,
where $N_1$ and $N_2$ are the number of atoms of  Na(1) and Na(2), respectively,
their spectra are resolved accordingly with the spectral weight of
$I_1$:$I_2$=$N_1$:$N_2$=1:2.
The NMR satellites are located at about 0.83~MHz (0.52~MHz) for Na(1) [Na(2)]
away from the center transition.

Upon cooling from 20~K to 8~K, all peaks move toward high frequencies
as shown in Fig.~\ref{fspec}(a), which indicates an increase of the Knight shift $K_n$
in the PM phase, with a positive hyperfine coupling constant among $^{23}$Na and Co$^{2+}$.
At even lower temperatures, the NMR spectra broaden significantly, for example, below
6~K at 1~T (Fig.~\ref{fspec}(a)) and below 5~K at 1.8~T(Fig.~\ref{fspec}(b)) ,
which clearly indicates the onset of magnetic ordering.
The satellites become indistinguishable due to broadening of the spectra
in the ordered phase. At low fields, with the change of local moment orientations
in the lattices, the relative orientation among the ordered moments and
the principle axis of the electric field gradient (EFG) also varies and
results in additional spectral broadening in the satellites.
At high fields, more NMR lines appear at low temperatures due to complicated
magnetic structures, which prevents us from distinguishing the center and the satellite
transitions. At a field of 2.3~T, as shown in Fig.~\ref{fspec}(c), all peaks are
resolvable again at low temperatures, consistent with
the spin polarized phase when magnetic ordering is suppressed by field~\cite{LiYuan_PRX_2022}.
Note that the integrated spectral weight, multiplied by temperature, remains as a constant at 1.9~T and above.
At low fields, the values drop by 30$\%$ below $T_N$ when cooled from the
PM phase to the ordered phase (data not shown).
We believe that quenched disorder broadens
the spectra and leads to partial loss of the signal.

To resolve the magnetic structures in the ordered phase,
the spectra are displayed at a low temperature
of 2~K with increasing fields, as shown in Fig.~\ref{fspec}(d).
Two broad NMR peaks are observed at 1~T, which is a direct evidence of AFM ordering.
With field increased from 1.4~T to 1.6~T, more peaks are resolved, where a
change of the magnetic structure is suggested.
For fields above 1.9~T, six NMR peaks are seen with overlaps of center and satellite
lines of $^{23}$Na from Na(1) and Na(2), where the fully polarized phase is achieved
as the ground state. Therefore, two magnetic phase transitions, at 1.4~T and 1.9~T
respectively, are revealed.

\begin{figure}[t]
\includegraphics[width=8.5cm]{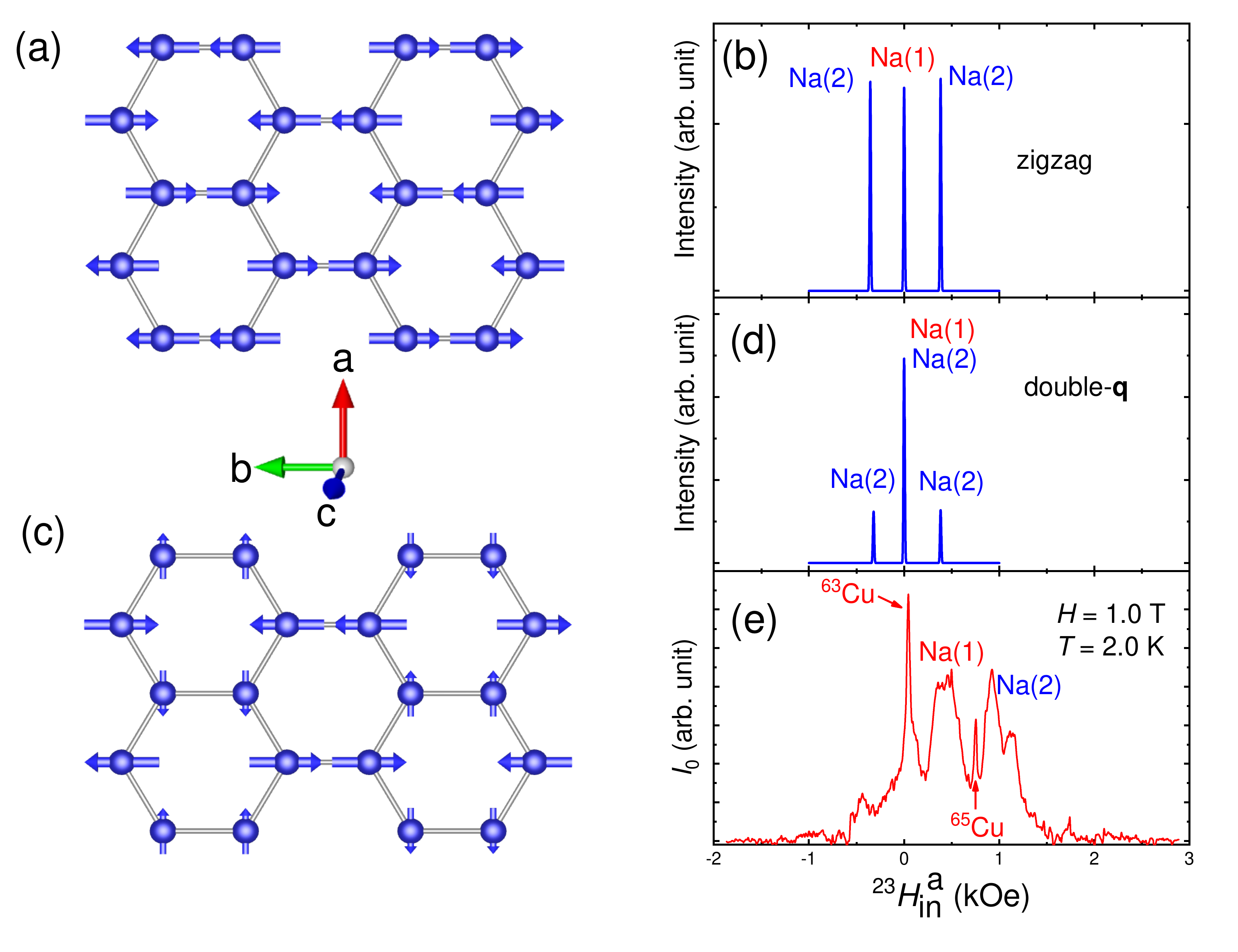}
\caption{\label{fdipolar}
{\bf Magnetic structures and simulated NMR spectra.}
(a)-(b) Spin pattern and center $^{23}$Na lines with
the zigzag order (see text).
(c)-(d) Spin pattern and center $^{23}$Na lines with
the double-{\bf q} order (see text).
(e) The measured NMR spectrum at 2~K at 1~T field.
}
\end{figure}

We simulated the low-field NMR spectra with two possible magnetic structures,
the zigzag order (Fig.~\ref{fdipolar}(a))~\cite{Yan_PRM_2019} and
the double-{\bf q} order (Fig.~\ref{fdipolar}(c))~\cite{GuYuchen_arXiv_2023}, respectively.
Since $^{23}$Na nuclei are distant from  Co$^{2+}$ ions,
only dipolar hyperfine couplings among them is taken into account.
For simplicity, the NMR satellites are not considered in the simulation.
The uniform magnetization of Co$^{2+}$ by the external field would induce a
rigid shift of the whole spectra, but does not affect the lineshape, therefore
not included in the calculations as well.

The total hyperfine field on each $^{23}$Na nucleus is then
obtained by summing over contributions from  ordered moments of Co$^{2+}$
with their relative coordinates.
The calculated center transition lines at low field are plotted in
Fig.~\ref{fdipolar}(b) and (d), with a local
moment of 0.9~$\mu_{\rm B}$/Co$^{2+}$ assumed for the zigzag phase~\cite{Yan_PRM_2019},
and two local moments of $\sqrt{3}$ and 1 (in unit of 0.9~$\mu_{\rm B}$/Co$^{2+}$)
in the double-{\bf q} phase~\cite{GuYuchen_arXiv_2023}, respectively.

For the zigzag pattern, Na(1) produces one center line at about 0~kOe,
and Na(2) produces two center lines at -0.326~kOe (-0.37~MHz) and 0.384~kOe (0.43~MHz),
with a relative spectral weight of 1:1:1 (Fig.~\ref{fdipolar}(b)).
By contrast, the double-{\bf q} pattern features one center line for Na(1) at 0~kOe,
and three center lines for Na(2) at -0.322~kOe (-0.36~MHz), 0~kOe, and 0.324~kOe (0.36~MHz),
which produces three lines with a relative weight of 1:4:1 (Fig.~\ref{fdipolar}(d)).

The simulated three-peak spectrum with equal weights for the zigzag pattern
is remarkably different from the actual spectrum with two peaks at 1~T (Fig.~\ref{fdipolar}(e))
which are assigned to be Na(1) and Na(2) respectively (see later).
On the other hand, for the double-{\bf q} order,
external field may have different effects on local moments in the non-collinear structure,
and the assumption of a rigid shift of spectra in the simulation may be invalid.
Therefore, our data do not support the zigzag order, but have no obvious contradiction
with the double-{\bf q} type.
We note that more distinguishable peaks emerge with fields above 1.1~T, which may
suggest a very weak incommensurate component superimposed
on the double-{\bf q} pattern and smear out the detailed spectra at fields of 1~T and below.

\section{{\label{skn} NMR Knight shift}}

The Knight shifts $^{23}K_n$ for both Na(1) and Na(2) at different fields are obtained from the
resonance frequency of the center peaks.
As shown in Fig.~\ref{fspec}(e), $^{23}K_n$ for Na(2) are shown as function
of temperatures. Using the relationship between Knight shift and the
reported susceptibility data~\cite{LiYuan_PRX_2022},
the hyperfine interaction $A_{hf}$ can be obtained by the slope of $^{23}K_n - \chi$ plot,
which is about 0.872~kOe/$\mu_{\rm B}$ and 1.166~kOe/$\mu_{\rm B}$ for Na(1) and Na(2) respectively,
as shown in the inset of Fig.~\ref{fspec}(f).
Note that the orbital contribution to $\chi$ is small and has been subtracted.
With this, the chemical shifts $K_C$, estimated by the $y-$intercept, are 0.017$\pm$0.005$\%$
and 0.032$\pm$0.006$\%$ for Na(1) and Na(2), respectively.

The spin contribution of Knight shift $K_s$ is then obtained as $K_s = K_n - K_C$.
In Fig.~\ref{fspec}(f), $^{23}K_s$  of both Na(1) and Na(2) are shown at 3~T,
where the rapid increase of $^{23}K_s$ upon cooling demonstrates PM behaviors.
Indeed, the high-temperature data of $^{23}K_s(T)$ follow the Curie-Weiss behavior,
$^{23}K_s = C/(T-\theta)$. 
$1/^{23}K_s$, demonstrated by functions of temperatures, 
are well fitted by straight lines with temperatures from 200~K down to 25~K.
With this, the Curie-Weiss temperature is obtained as $\theta$ = 1.1$\pm$0.3~K
for both Na(1) and Na(2).

\begin{table}[htbp]
   \centering
   \caption{Curie-Weiss temperature $\theta$ obtained from different measurements with field applied along the $a$ axis.}
   \label{tab:1}
   \begin{tabular}{cccc}
      \hline\hline\noalign{\smallskip}
       Measurements ~ & ~$^{23}K_n$ [Na(1)]~ & ~$^{23}K_n$ [Na(2)]~ &~ $M$ [Low-T]~\cite{LiYuan_PRX_2022}~ \\
      \noalign{\smallskip}\hline\noalign{\smallskip}
       $\theta$ (K) & 1.1$\pm$0.3 & 1.1$\pm$0.3 & 1.0 \\
      \noalign{\smallskip}\hline
   \end{tabular}
\end{table}

By comparison, the magnetic susceptibility data also reveal
a positive $\theta$ = 1.0~K, shown in Table~\ref{tab:1},
by data fitting from 20~K to 120~K with the same field orientation~\cite{LiYuan_PRX_2022}.
These small but positive Curie-Weiss temperatures obtained at high temperatures
indicate that the material contains FM intralayer couplings.
Given that the ground state has an AFM order~\cite{Wong_JSSC_2016,Yan_PRM_2019},
highly competing magnetic FM and AFM interactions are expected
with such a small $\theta$. Then our data support the existence of
FM interactions as proposed for cobaltates with honeycomb lattices~\cite{Liu_PRB_2018},
although the absolute values of exchange couplings remain to be determined.

\section{{\label{sslrr} Spin-lattice relaxation rates and magnetic transitions}}

\begin{figure}[t]
\includegraphics[width=8.5cm]{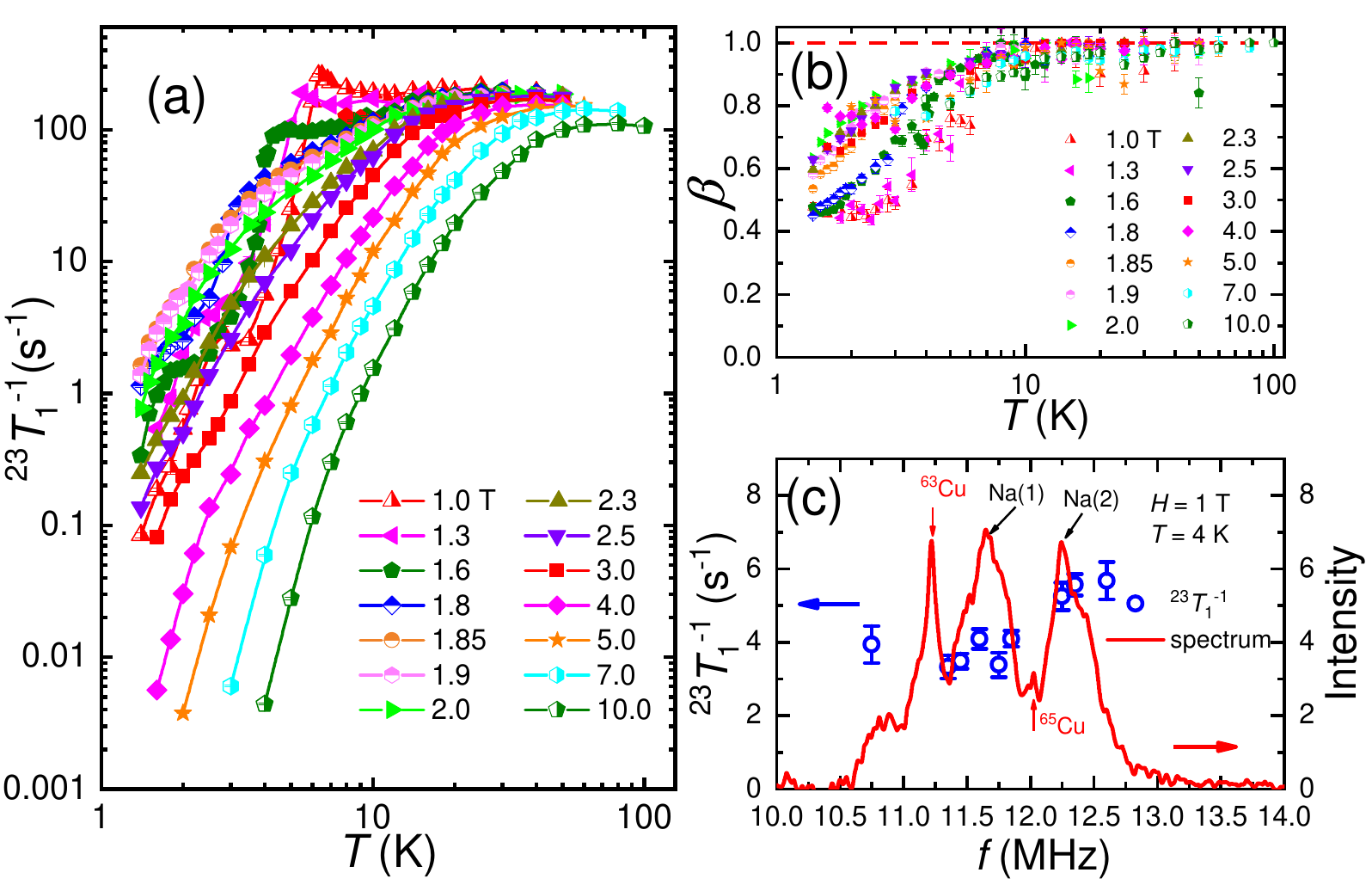}
\caption{\label{ft1}
{\bf Spin-lattice relaxation rates.}
(a) $1/^{23}T_1$ of Na(2) as functions of temperatures measured under typical fields.
(b) The stretching factor $\beta$ as functions of temperatures.
(c) $1/^{23}T_1$ (left scale) measured on different frequencies of the spectrum (right scale).
}
\end{figure}

$1/^{23}T_1$ was measured to determine the low-energy spin fluctuations.
$1/T_1$ measures the low-energy spin dynamics with
$1/T_1=T\Sigma_q{A_{hf}^2(q)\frac{Im\chi(q,\omega)}{\omega}}$,
where $A_{hf}(q)$ is the hyperfine coupling constant,
$\chi$ is the dynamic susceptibility of electrons,
$q$ is the wave vector, and $\omega$ is the NMR frequency.
$1/^{23}T_1$ was measured at various frequencies across
the spectra line at 4~K and 1~T, as shown in Fig.~\ref{ft1}(c).
$1/^{23}T_1$ on the right peak is about 50\% larger than
that on the left peak, which supports assignment of the right peak
to Na(2), because Na(2) has a stronger hyperfine coupling as revealed before.

Here we primarily report $1/^{23}T_1$ measured on the center peak of Na(2).
As shown in Fig.~\ref{ft1}(a), $1/^{23}T_1$ is displayed as functions of
temperatures, with fields from 1~T up to ~10 T.
The stretching factor $\beta$~$\approx$~1 (Fig.~\ref{ft1}(b))
in the PM phase, indicates high quality of the sample.
Upon cooling, a rapid decrease of $\beta$ is seen in the ordered phase
and also in the fully polarized phases, which coincides with the
spectral broadening at low temperatures (Fig.~\ref{fspec}(a)-(c)).
We think that the spectral broadening and decrease of $\beta$ are
caused by quenched disorder which is very effective in the ordered phase and fully
polarized phase when the local moments are large.

At low fields, $1/^{23}T_1$ increases slightly upon cooling through 100~K,
before entering the ordered phase, which evidences the onset of low-energy
spin fluctuations. In the following,
three types of magnetic transitions are revealed.

\begin{figure}
\includegraphics[width=8.5cm]{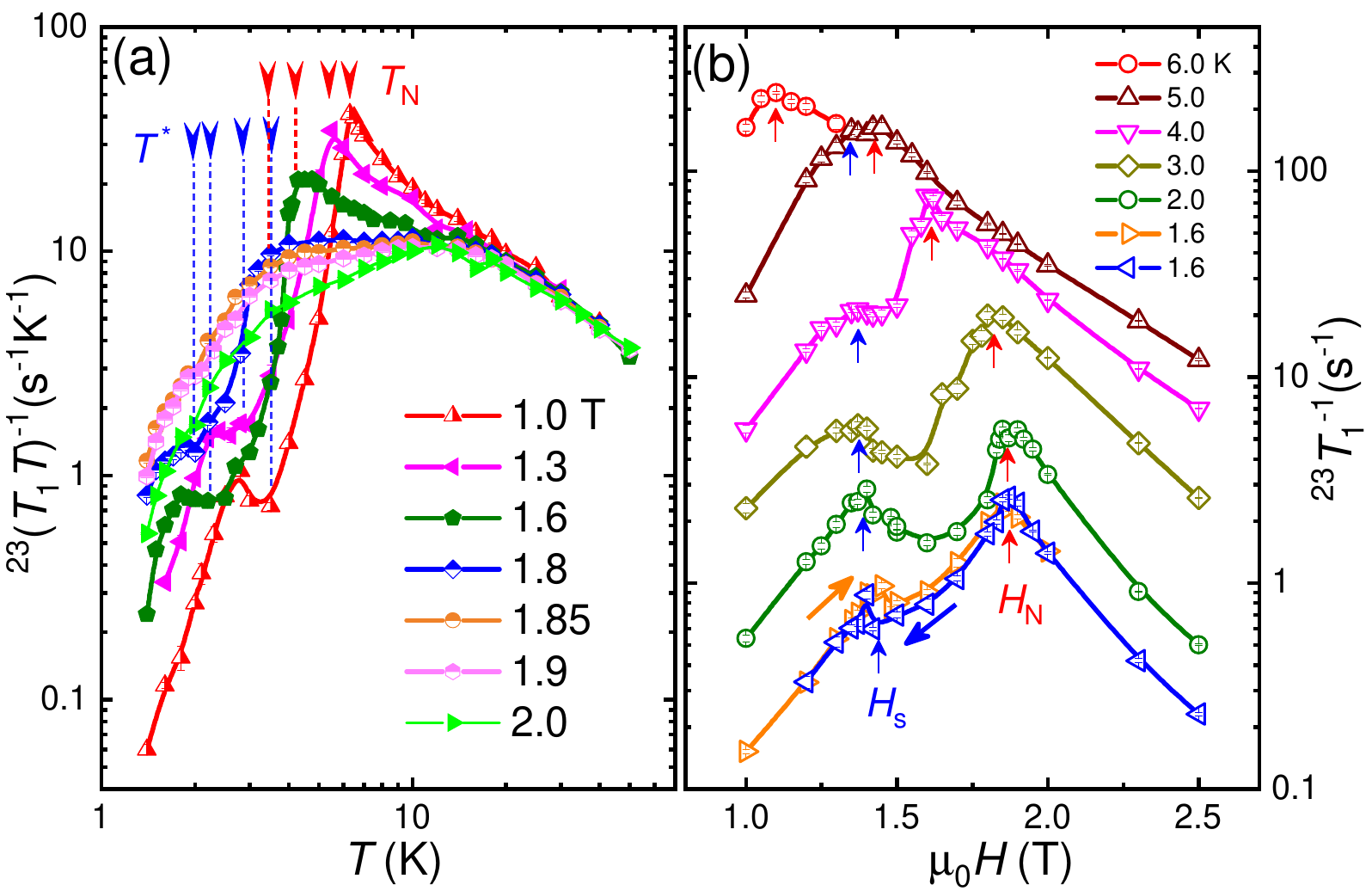}
\caption{\label{ftn}
{\bf Determination of magnetic phase transitions.}
(a) $1/^{23}T_1T$ as functions of temperatures at low fields.
Red and blue arrows mark the peak and ``knee"-like feature
in the data, denoting the double transition temperatures of $T_{\rm N}$ and $T^ *$, respectively.
(b) $1/^{23}T_1$ as functions of fields under constant temperatures from 1.6~K to 6~K.
Up arrows denote the fields of two magnetic transitions occurring at $H_{\rm N}$ and $H_{\rm s}$ respectively.
Left and right arrows at 1.6~K illustrate the field ramping directions, where magnetic hysteresis
is seen at about $H_{\rm s}$ with opposite ramping directions.
}
\end{figure}

First, the spin-lattice relaxation rate divided by temperature,
as shown in Fig.~\ref{ftn}, reveals all
the N\'{e}el transitions at low fields, characterized by
a peaked feature in $1/^{23}T_1T$ at $T_{\rm N}$.
$T_{\rm N}$ is about 6.3~K at 1~T, 
and barely resolvable at about 3.5~K with field at 1.8~T, 
below which $1/^{23}T_1T$ drops sharply with a gapped behavior.
With fields from 1.85~T to 1.9~T, the peaked behavior in the $1/^{23}T_1T$
is not resolvable, and the phase boundary will be determined
later by a field-sweep measurement on $1/^{23}T_1T$. At 2~T and above,
$1/^{23}T_1T$ exhibits a smooth decrease with temperature, which indicates
the suppression of the AFM ordering.
The detailed $T_{\rm N}$ at different fields are then added in the phase
diagram (Fig.~\ref{fqcr}). The suppression of $T_{\rm N}$
is consistent with earlier susceptibility measurements~\cite{Wong_JSSC_2016, Yan_PRM_2019};
however, our values of $T_{\rm N}$ are slightly higher, which is probably
caused either by a shorter time scale of $T_1$ measurements
or a small difference in the field alignment.

Second, with fields from 1~T to 1.8~T, a ``knee"-like
feature is seen in $1/^{23}T_1T$ at temperatures far below $T_{\rm N}$ (
Fig.~\ref{ftn}(a)), with the onset temperatures marked as $T^*$.
This ``knee"-like behavior, rather than a continuous gapped behavior
below $T_{\rm N}$ as expected with an in-plane field, reveals
emergence of additional enhanced low-energy spin fluctuations.
$T^*$ at different fields are then determined and shown in
Fig.~\ref{fqcr}, which is about 3.5~K at 1~T, decreases to 2~K
at 1.8~T, and diminishes at about 1.85~T. Therefore,
$T^*$ follows the same trend as $T_{\rm N}$ with field.
By this, we think that a weak spin reorientation may occur
in the ordered phase~\cite{Lee_PRB_2021}.  However,
such transition is not seen in the spectra, which may be
masked by the broad spectra if the change of the magnetic
structure is not large.

Third, $1/^{23}T_1$, measured as functions of fields at 6~K and below,
demonstrates a double-peak feature with field, as shown in Fig.~\ref{ftn}(b).
The field values of the peak position in  $1/^{23}T_1$
are labeled as $H_{\rm N}$ and $H_{\rm s}$, respectively.
Each $H_{\rm N}$ with the corresponding temperature is consistent
with $T_{\rm N}$  measured at the same field, which therefore
represents the N\'{e}el ordering as discussed before.
At 1.6~K, $H_{\rm N}\approx$~1.87~T; by extrapolation,
the ``critical" field $H_{\rm C}\approx$1.9~T at zero temperature.

\begin{figure}[t]
\includegraphics[width=8.5cm]{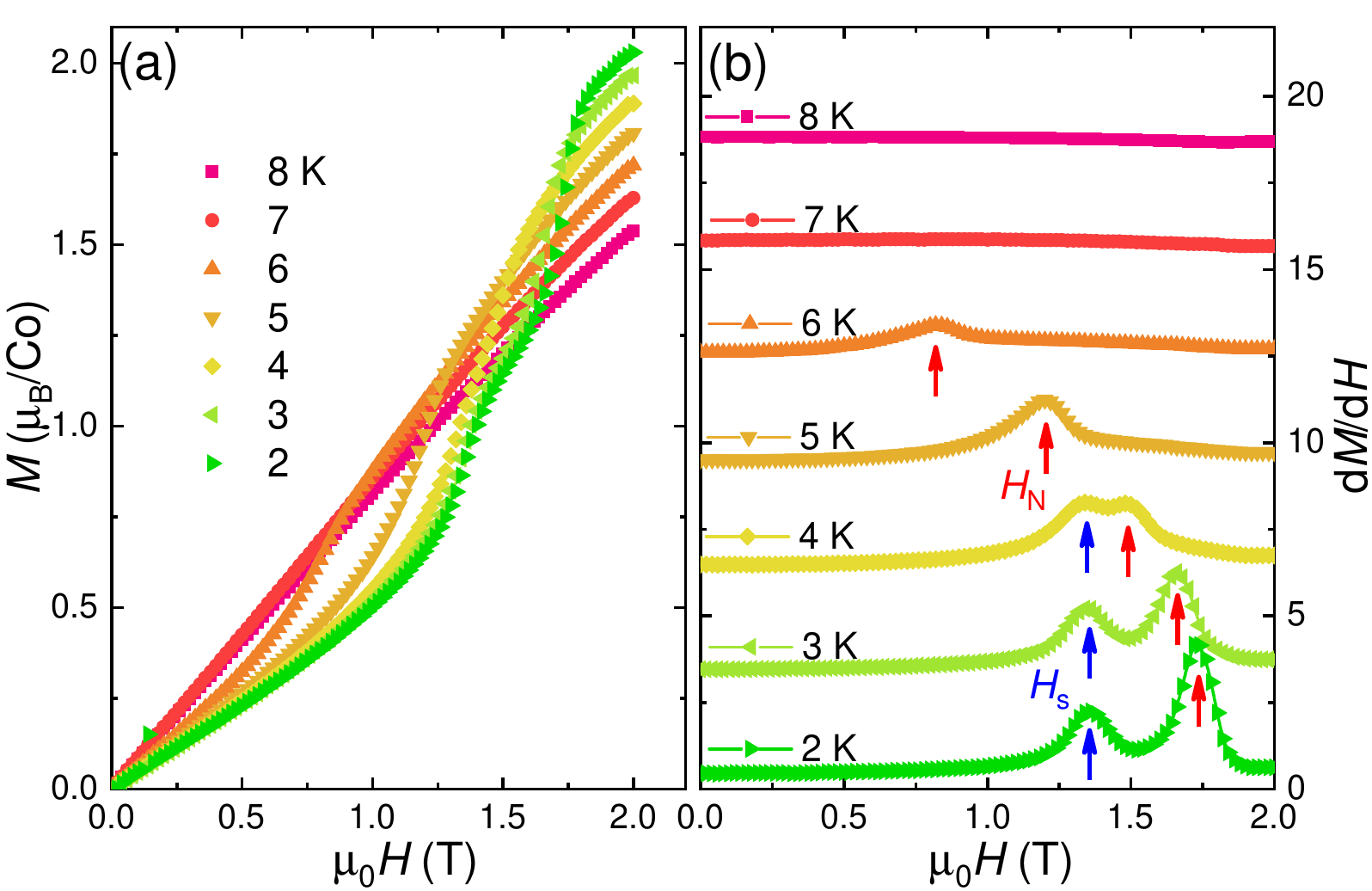}
\caption{\label{fmh}
{\bf Magnetization data of the sample.}
  (a) $M(H)$ as functions of fields measured at selected temperatures.
  (b) Calculated $dM(H)/dH$, at different temperatures.
   Red and blue arrows mark the peak positions of the data, labeled
   as $H_{\rm N}$ and $H_{\rm s}$, respectively. Data are shifted vertically for clarity.
}
\end{figure}

On the other hand, $H_{\rm s}$ remains at a constant field
of $\sim$1.4~T, but always lower than $H_{\rm N}$.
At 1.6~K, a hysteresis behavior close to $H_{\rm s}$ is also revealed,
by the data difference with opposite field ramping directions
as marked in Fig.~\ref{ftn}(b), where $H_{\rm s}$ tends
to be lower with the field ramped down.
Far above or below $H_{\rm s}$, the hysteresis
behavior diminishes. This low-field peak and its
hysteresis should indicate an additional first-order magnetic transition.

Such a transition at a constant field is further verified by our
$dc$ magnetization measurements. As shown in Fig.~\ref{fmh}(a),
the low-temperature magnetization $M(H)$ data are shown with
field up to 2~T at selected temperatures from 8~K down to 2~K.
At 2~K, double magnetic transitions are clearly seen by the change
of the slop in $M(H)$. For accuracy, the derivative $dM/dH$ is then
calculated and plotted in Fig.~\ref{fmh}(b).
At the temperatures of 2, 3, and 4~K, $dM/dH$ clearly shows the double-peak
feature, with fields also labeled as $H_{\rm N}$ and $H_{\rm s}$, respectively.
Again, $H_{\rm s}$ remains at about 1.4~T and barely changes with temperature.

For comparison, the $H_{\rm N}$ and $H_{\rm s}$ determined by $1/^{23}T_1$
and by $dM/dH$ at each temperature are added in the phase diagram in
Fig.~\ref{fqcr}. It is clearly seen that both measurements
are consistent. The small difference in the field values extracted from
the measurements may be due to field-calibration errors and/or a small tilting
of the field towards the ``harder" $a$-axis in the $1/^{23}T_1$ measurement.
In fact, the transition at $H_{\rm s}$ is consistent with the reported
first-order phase transition from an AFM ordering to a ``1/3-AFM" ordering,
resolved by the neutron diffraction~\cite{LiYuan_PRX_2022}.
The AFM wave vectors switches from $(\pm a^*/2,\pm b^*/2,0)$ to
$(\pm a^*/3,\pm b^*/3,\pm c^*/3)$ through the transition, where
the magnetic cell is enlarged.

\section{{\label{ssg} Spin gaps close to the critical field}}

In principle, a quantum disordered phase is expected for fields above
$H_{\rm C}$, where the spin gap increases monotonically.
However, as we show below, the low-temperature spin gap
may exhibit a non-monotonic field dependence in this region, which
suggests an additional phase.

\begin{figure}
\includegraphics[width=8.5cm]{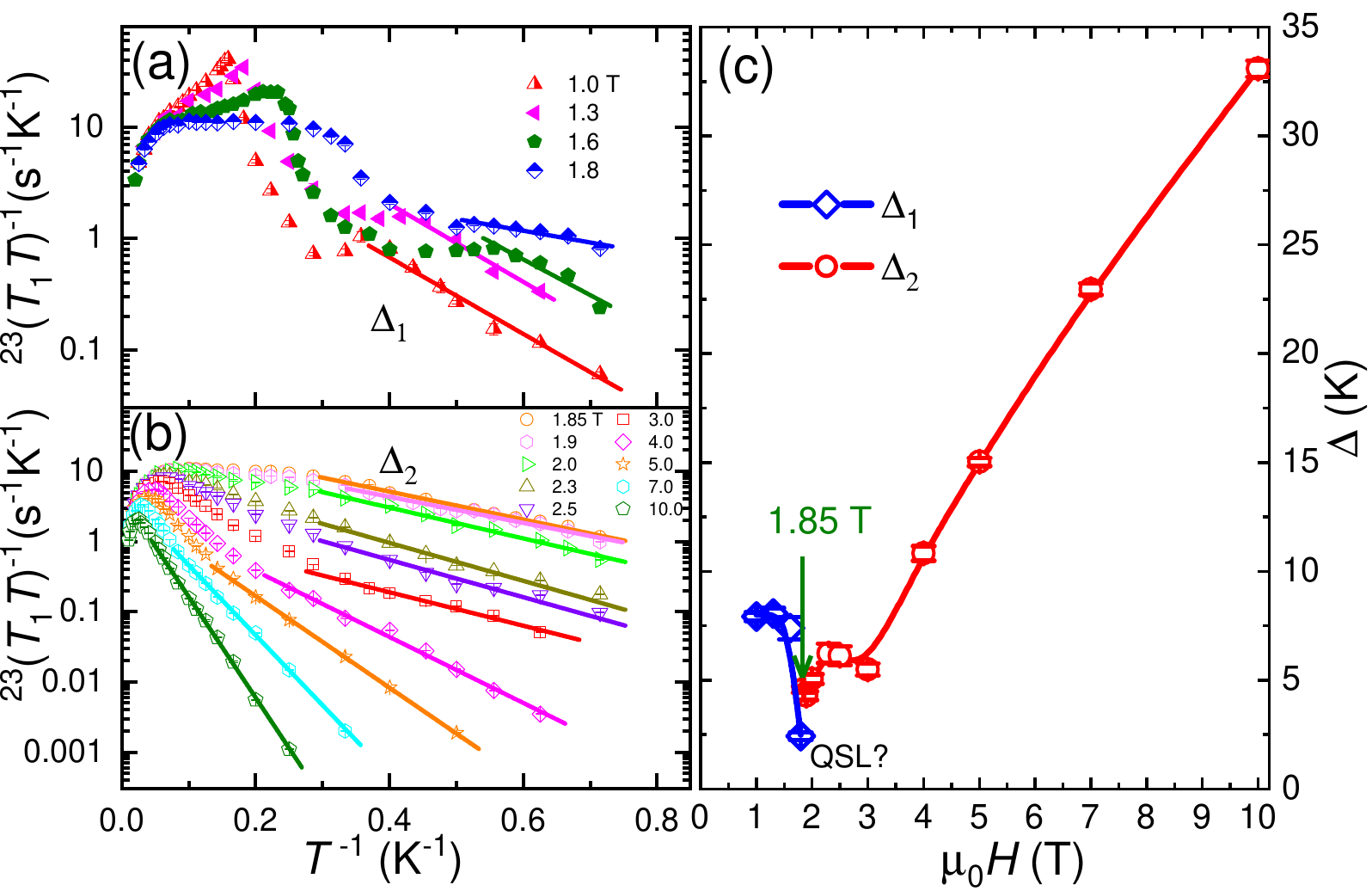}
\caption{\label{ftitgap}
{\bf Low-temperature spin gaps.}
(a)-(b) The semi-log plots of $1/^{23}T_1T$ as functions of $1/T$,
below and above the ``critical" field (see text), respectively.
The straight lines are linear fits to data in the low-temperature regime,
which determine the gaps $\Delta_1$ and $\Delta_2$ as labeled.
(c) $\Delta_1$ (in the ordered phase) and $\Delta_2$ (in the disordered phase),
as functions of fields. Solid lines are guides to the eyes. The
down arrow marks the transition field at about 1.85~T.
}
\end{figure}

$1/^{23}T_1$ exhibits a downturn behavior
at low temperatures, which suggests the onset of spin gap both in the
low-field ordered phase and the high-field spin polarized phase.
By which, we fit $1/^{23}T_1T$ to an empirical function,
$1/T_1T \propto e^{-{\Delta/T}}$.
In the ordered AFM phase, $1/^{23}T_1T$ as functions of $1/T$,
as shown by the semi-log plots in Fig.~\ref{ftitgap}(a),
follow straight lines in the low temperature regime.
Then the gaps, defined as $\Delta_1$,  are obtained by
the linear fit and shown in Fig.~\ref{ftitgap}(c) at
different fields.
The onset of the gapped behavior in the ordered phase, rather than
a power-law temperature dependence of $1/T_1(T)$, suggests an
ordering with local moments at least partially along
the $a$ axis, where a gap opens due to longitudinal fields.
Similarly, above $H_{\rm C}$,
a gapped behavior is also followed as shown
in Fig.~\ref{ftitgap}(b), and the obtained gap values,
$\Delta_2$, are also added in Fig.~\ref{ftitgap}(c).

Interestingly, the gap does not drop to zero at the boundary of the ordered phase
and the disordered phase.
$\Delta_1$ decreases
with field and $\Delta_2$ increases with field.
They cross at fields between 1.8 and 1.9~T;
however, none of them reach zero from field between 1.8~T and 1.9~T.
The presence of gap is also directly demonstrated by the sharp
downturns in the log-log plot of $1/^{23}T_1T$ in all of the fields as
shown in Fig.~\ref{ftn}(a).
The absence of gapless excitations rules out a quantum critical
point in the current system; instead, a weakly
first-order quantum phase transition is suggested with the onset
of a small gap at the transition field.

Furthermore, $\Delta_2$ demonstrates a non-monotonic field
dependence as shown in Fig.~\ref{ftitgap}(c):
it increases first with field and then decreases;
above 3~T, a large increase emerges again as expected for
a fully polarized phase. With fields from 1.85~T to 3~T,
a dome-like behavior is observed.
Such a dome-like shape of $\Delta_2$ strongly supports
the existence of an additional phase
between the AFM phase and the fully polarized phase.

\section{{\label{sqcr} Quantum criticality and phase diagram}}

\begin{figure}[t]
\includegraphics[width=8.5cm]{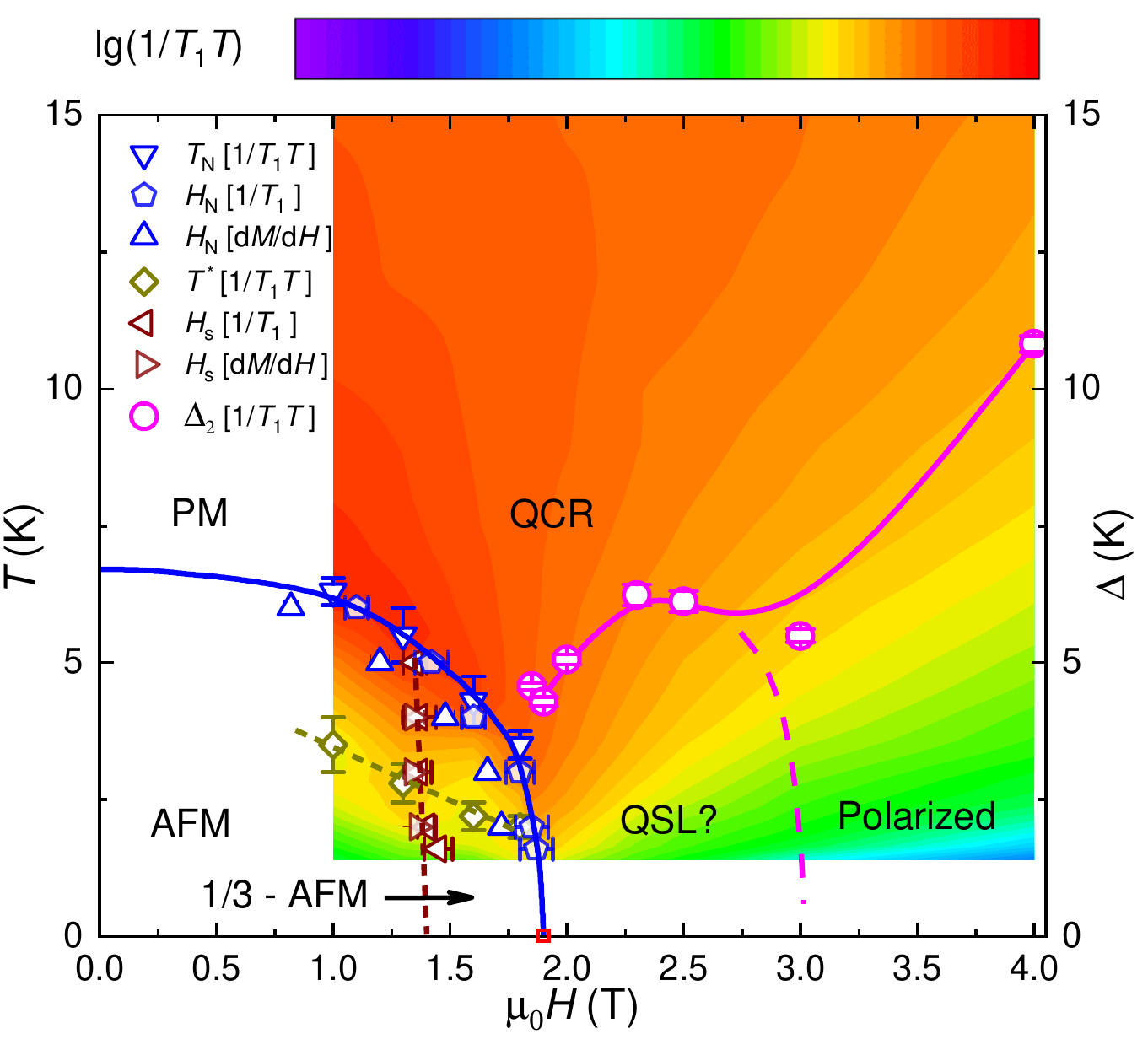}
\caption{\label{fqcr}
{\bf Phase diagram.}
Colored map represents the contour plot of $1/^{23}T_1T$ data,
with a QCR at high temperatures.
$T_{\rm N}$, $H_{\rm N}$, $T^*$, $H_{\rm s}$, and $H_{\rm C}$
are phase boundaries (see text) determined by different
measurements as labeled. $\Delta_2$ is the high-field spin gap.
Lines are guides to the eyes.
}
\end{figure}

A colored contour plot of $1/^{23}T_1T$ is shown
in Fig.~\ref{fqcr} as a function of ($H$, $T$).
$1/^{23}T_1T$ is maximized when the temperature is close to
$T_N$, which is a typical feature for a renormalized
classical regime.
Below  2~K, $1/^{23}T_1T$ is maximized at $H_{\rm C} {\approx}$1.9~T,
which is a signature of quantum phase transition with strong quantum fluctuations.
At high temperatures, a quantum critical region (QCR),
as noted in Fig.~\ref{fqcr}, is also seen as a funnel shape in the color map, when
it is getting close to 1.9~T.

With detailed data analysis performed in Sec.~\ref{sslrr},
a complete magnetic phase diagram is established as
shown in Fig.~\ref{fqcr}, where the PM, AFM,
1/3-AFM, disordered phases and their phase boundaries
are determined. The gap in the disordered phase, $\Delta_2$,
is also added in the phase diagram. A QSL region
is speculated just between the ordered phase and the fully
polarized phase where $\Delta_2$ exhibits a dome-like
shape with field.

We also compared our results with a previous NMR work on
polycrystals, where the AFM phase,
the potential QSL phase, and the fully polarized phase are
also proposed~\cite{Vavilova_PRB_2023}. The usage of single-domain single crystal allows us to
resolve all the phase boundaries precisely. In addition,
we identified a possible spin reorientation transition below $T_{\rm N}$
and a first-order phase transition at 1.4~T.
The reported $1/^{23}T_1(T)$ of polycrystals exhibits a two-gap feature in a large field
range~\cite{Vavilova_PRB_2023}, which is absent in our work and therefore should be
attributed to the magnetic anisotropy of the material.

\section{{\label{sd} Discussions}}

The complicated phase diagram indicates more
competing exchange interactions in this system, which needs to be
further studied by different probes.
In particular, the nature of the transition at $T^*$,
remains unknown and needs to be verified by other measurements.
One candidate mechanism for the transition is the DM interaction, which
could be effective far below $T_{\rm N}$. Given that
the inversion symmetry is broken among next-nearest neighboring Co$^{2+}$
ions in the honeycomb lattice, a weak DM interaction is possible~\cite{Chen_PRX_2018}.
Because NMR is very sensitive to the low-energy spin fluctuations,
such weak DM interaction may lead to the observations below $T^*$.
However, we are cautious that DM
interaction has not been reported by other studies~\cite{Wong_JSSC_2016, Yan_PRM_2019}.

Theoretically, a first-order phase transition may occur if a QSL exists
between the ordered phase and the spin-polarized
phase~\cite{Makhfudz_PRB_2014,Rico_PRB_2020,Tomishige_PRB_2018}.
We speculate that Kitaev couplings ($K$-term) and off-diagonal ($\varGamma$ term)
may exist and strongly affect the spin dynamics in the system~\cite{Luo_PRB_2021,Luo_QM_2021}.
For comparison, in two other Kitaev materials, $\alpha$-RuCl$_3$ and
Na$_2$Co$_2$TeO$_6$, a new phase seems to be established
between the ordered phase and the polarized phase under in-plane magnetic field,
where a QSL has been suggested~\cite{YJKim_PRB_2017, Zheng_PRL_2017,Lin_NC_2021}.
A gapless behavior is observed in the low-temperature $1/T_1$ data of $\alpha$-RuCl$_3$,
which supports a proximate Kitaev QSL~\cite{Zheng_PRL_2017}.
For the current compounds, we think that a QSL may also exist, given
the existence of a dome-shape of $\Delta_2$.

\begin{figure}[t]
\includegraphics[width=8.5cm]{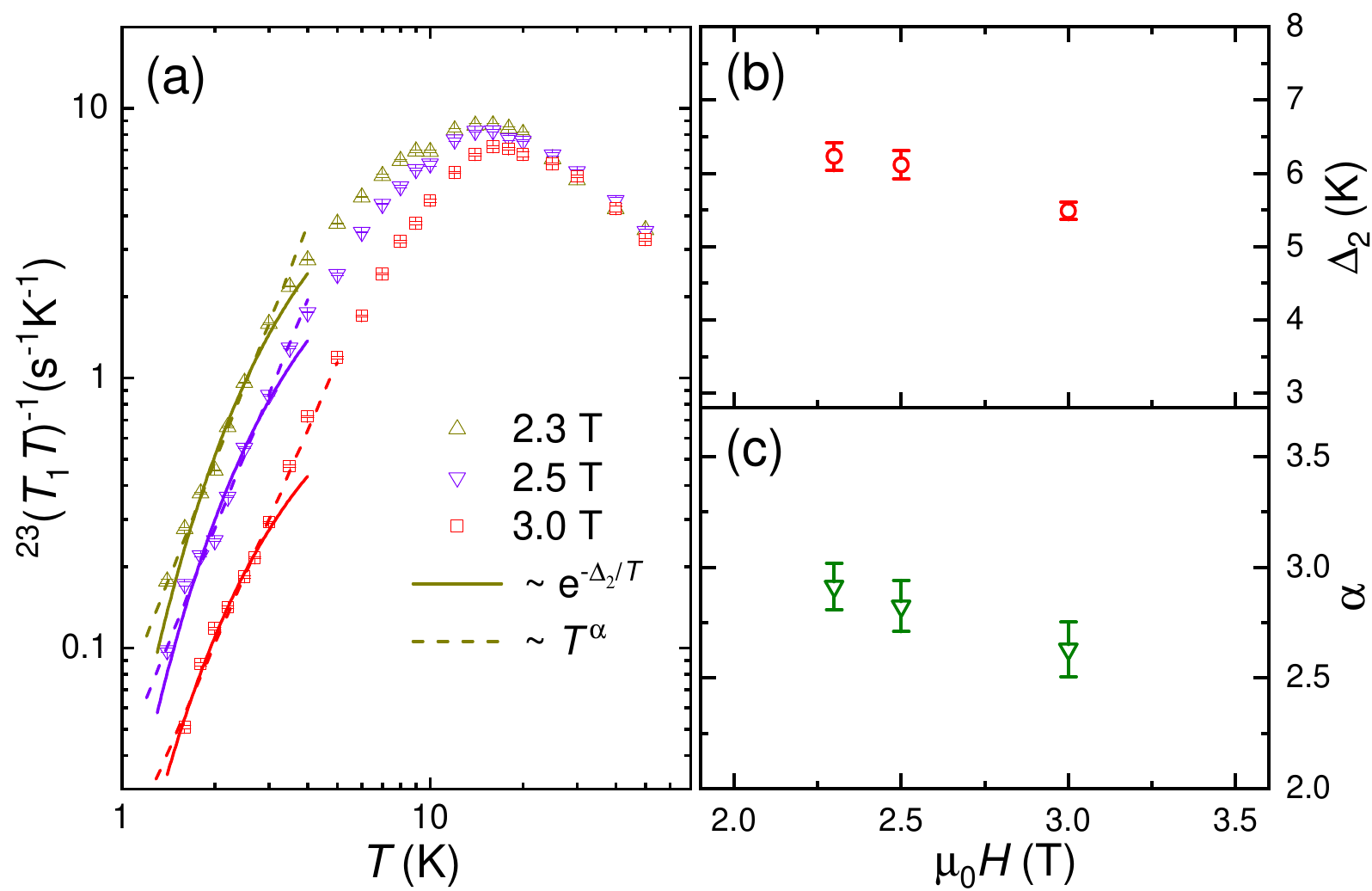}
\caption{{\bf Comparison of gap and power-law fits to $1/^{23}T_1T$.}
(a) $1/^{23}T_1T$ as functions of temperatures at 2.3, 2.5 and 3.0~T.
The solid lines and dash lines represent the fit to the gapped function and the power-law function
below 3~K, respectively. (b) $\Delta_2$ as a function of field.
(c) Power-law exponent $\alpha$ as a function of field.}
\label{fig:gpl}
\end{figure}

However, we found that a power-law fitting is also applicable
to the low temperature data  $1/^{23}T_1T$ with field just above $H_{\rm C}$.
As shown in Fig.~\ref{fig:gpl}(a), function fit to either a gapped
behavior and a power-law behavior in the same temperature range
is performed at temperatures below 3~K, with fields from 2.3 to 3~T.
The obtained gap $\Delta_2$ and the power-law exponent $\alpha$
are depicted in Fig.~\ref{fig:gpl}(b) and (c), respectively,
as function of field. Notably, both $\Delta_2$ and $\alpha$
decrease with increasing field, contradicting to the expected
increase of both quantities with field in the fully polarized phase.
Such anomalous behavior may support a QSL
intervals between the ordered phase and the fully polarized phase, 
although we cannot differentiate a gapped or a gapless behavior
with current data.
\\
\\

\section{{\label{ssum}}Summary and Acknowledgement}

In summary, we investigated the static and low-energy dynamical behavior
through NMR experiments on a high-quality Na$_3$Co$_2$SbO$_6$
single crystal. With field applied along the $a$ axis, our data
reveal a positive Curie-Weiss constant at high temperatures, which supports
the existence of FM exchange couplings.
Given the absence of FM ordering, such a FM
coupling may not be the Heisenberg type.
The observation of three separate transition lines with field, including
$T_N$, $T^*$ and $H_{\rm s}$, further suggests complex magnetic exchange couplings
in the system, which may help to establish QSLs.
Indeed, despite our observation of a QCR
at high temperatures, the low-temperature gap in the magnetically
disordered phase shows a non-monotonic field dependence, which may
be a signature of QSL. We hope that inelastic neutron scattering
may help to address this by looking for excitation continuum.

The authors thank Zhengxin Liu, Zhiyuan Xie,
Jie Ma, and Yuan Wan for helpful discussions. This work is supported by
the National Key R\&D Program of China (Grant Nos. 2023YFA1406500, 2022YFA1402700
and 2021YFA1401900) and the National Natural Science Foundation
of China (Grants Nos. 12134020, 12374156 and 12061131004)

%\bibliographystyle{apsrev4-1}
%\bibliography{NCSO}

%

\onecolumngrid

%\newpage

\end{document}